\documentclass[preprint, twocolumn, 5p]{elsarticle}
\usepackage{lmodern}
\usepackage{amssymb,amsmath}
\usepackage{ifxetex,ifluatex}
\usepackage{fixltx2e} 
\ifnum 0\ifxetex 1\fi\ifluatex 1\fi=0 
  \usepackage[T1]{fontenc}
  \usepackage[utf8]{inputenc}
\else 
  \ifxetex
    \usepackage{mathspec}
  \else
    \usepackage{fontspec}
  \fi
  \defaultfontfeatures{Ligatures=TeX,Scale=MatchLowercase}
\fi
\IfFileExists{upquote.sty}{\usepackage{upquote}}{}
\IfFileExists{microtype.sty}{%
\usepackage{microtype}
\UseMicrotypeSet[protrusion]{basicmath} 
}{}
\usepackage{natbib}
\ifx\paragraph\undefined\else
\let\oldparagraph\paragraph
\renewcommand{\paragraph}[1]{\oldparagraph{#1}\mbox{}}
\fi
\ifx\subparagraph\undefined\else
\let\oldsubparagraph\subparagraph
\renewcommand{\subparagraph}[1]{\oldsubparagraph{#1}\mbox{}}
\fi
\usepackage{booktabs}

\date{}

\begin{document}

\begin{frontmatter}
\title{First-principles Study of Rashba Effect in Ultra-thin Bismuth Surface Alloys}

\author[NST]{Naoya Yamaguchi}
\ead{n-yamaguchi@cphys.s.kanazawa-u.ac.jp}
\author[NST,ESICB]{Hiroki Kotaka}
\author[ISE]{Fumiyuki Ishii}
\ead{ishii@cphys.s.kanazawa-u.ac.jp}
\address[NST]{Division of Mathematical and Physical Sciences, Graduate School of Natural Science and Technology, Kanazawa University, Kanazawa 920-1192, Japan.}
\address[ESICB]{Elements Strategy Initiative for Catalysts and Batteries (ESICB),
Kyoto University, Kyoto 615-8245, Japan.}
\address[ISE]{Faculty of Mathematics and Physics, Institute of Science and Engineering, Kanazawa University, Kanazawa 920-1192, Japan.}
\begin{abstract}
We performed density functional calculations for
ultra-thin bismuth surface alloys: surface alloys of bismuth and face-centered cubic metals Bi/\emph{M}(111)-($\sqrt{3}\times\sqrt{3}$)\emph{R}30° (\emph{M}=Cu, Ag, Au, Ni, Co, and Fe).
Our calculated
Rashba parameters for the Bi/Ag are consistent with the previous experimental
and theoretical results. We predicted a trend in the Rashba coefficients \(\alpha_R\) of
bands around the Fermi energy for noble metals as follows: Bi/Ag $>$ Bi/Cu $>$ Bi/Au.
As for the transition metals, there is a trend in \(\alpha_R\): Bi/Ni
$>$ Bi/Co $>$ Bi/Fe.
Our finding may lead to design efficient
spin-charge conversion materials.
\end{abstract}
\end{frontmatter}

\section{Introduction}\label{introduction}

Strong spin-orbit interaction (SOI) is originated from a heavy element
such as bismuth or spatial inversion symmetry breaking, that is,
electric field. Rashba-Bychkov effect is induced by SOI
\citep{Bychkov_Properties_1984, LaShell_Spin_1996} and mainly important
to spintronics devices such as spin FET (field effect transistor)
\citep{Datta_Electronic_1990}. The symmetry breaking can take place for
the two dimensional electronic gas (2DEG) at the surfaces and
interfaces. SOI is attracted attention due to spintronics applications
in recent years, and Rashba effect is extensively studied
\citep{Ishizaka_Giant_2011}.

Spin-to-charge conversion, conversion phenomenon between spin and charge
currents, is one of the most important topics in spintronics. Spin
current is generated by spin Seebeck effect (SSE)
\citep{Uchida_Observation_2008}, for example. SSE can transform
temperature gradient into spin current besides charge current which
originate from normal Seebeck effect. Spin-to-charge conversion is
important to convert spin current into charge current so effectively as
to increase thermoelectric conversion efficiency using the spin Seebeck
effect. Inverse spin hall effect (ISHE) is known as such a conversion
mechanism \citep{Saitoh_Conversion_2006}. Recently, for a Bi/Ag Rashba
interface, a large conversion phenomenon between spin and charge
currents has been reported \citep{Sanchez_Spin_2013}. That
spin-to-charge conversion is induced by the inverse Rashba-Edelstein
effect (IREE) on the metal-metal or metal-insulator interface with
strong SOI, which is the inverse process of Edelstein effect
\citep{Edelstein_Spin_1990}. The strength of the IREE is related with
the Rashba coeffecient \(\alpha_R\), which reflects the strength of the
Rashba effect. IREE can be used for an alternative mechanism of
detection of spin current utilizing ISHE. Indeed, giant Rashba spin
splitting in Bi/Ag surface alloys was reported
\citep{Ast_Giant_2007, Bian_Origin_2013}.

In this paper, we investigated the Rashba effects for surface alloys
composed of Bi and 3d-transition metals (Fe, Co, Ni) and noble metals
(Cu, Ag, Au) using first-principles calculations. The trend in the
surface alloys can be considered to be similar with that in the
corresponding interfaces. One of our results for noble metals is that
there is a trend in the order of magnitude of \(\alpha_R\) for bands around Fermi energy: Bi/Ag
\textgreater{} Bi/Cu \textgreater{} Bi/Au.

\section{Methods}\label{methods}

For the two dimensional electronic system perpendicular to the direction
of electric field, the Hamiltonian to be considered is the sum of the
kinetic energy of free electron gas and Rashba Hamiltonian which can be
expressed by
\[ H_R=\alpha_R(\hat{e}_{z}\times\vec{k}_{||})\cdot\vec{\sigma}=\alpha_R(k_y\sigma_x-k_x\sigma_y), \]
where \(\alpha_R\) denotes the Rashba coefficent, \(\hat{e}_{z}\) the
unit vector along \(z\) axis, \(\vec{k}_{||}=(k_x, k_y, 0)\) the wave
vector, and \(\vec{\sigma}=(\sigma_x, \sigma_y, \sigma_z)\) the Pauli
matrices vector. We can solve the eigen value problem for
that Hamiltonian, and obtain the energy dispersion relation for the
nearly free electron under the Rashba effect:
\[ E_\pm(\vec{k}_{||})=\frac{\hbar^2k_{||}^2}{2m^*}\pm\alpha_R k_{||}, \]
where \(\hbar\) is the Planck constant, and \(m^*\) the effective
mass of electrons. Then, the Rashba coefficient can be obtained as
follows: \(\alpha_R=2E_R/k_R\) where \(E_R=m^*\alpha_R^2/(2\hbar^2)\) is
the Rashba energy, and \(k_R=m^*\alpha_R/\hbar^2\) the Rashba momentum
offset.

We performed density functional calculations within the local spin density approximation \citep{Ceperley_Ground_1980} using OpenMX code \citep{Ozaki_OpenMX}, with the fully relativistic total angular momentum dependent pseudopotentials taking SOI into account \citep{Theurich_Self_2001}. We
adopted norm-conserving pseudopotentials with an
energy cutoff of 200 Ry for charge density including 5\emph{d}, 6\emph{s}, and 6\emph{p}-states as valence states for Bi; the 3\emph{s},
3\emph{p}, 3\emph{d}, and 4\emph{s} for Cu, Ni, Co, and Fe; 4\emph{p}, 4\emph{d}, and 5\emph{s} for Ag; 5\emph{p}, 5\emph{d}, and 6\emph{s} for Au.
We used \(8\times8\times1\)
k-point mesh. In this calculation, the numerical pseudo atomic orbitals
are utilized as follows: For all models, the numbers of the \emph{s}-, \emph{p}-, and
\emph{d}-character orbitals are three, three, and two, respectively; The cutoff
radiuses of Bi, Cu, Ag, Au, Ni, Co, and Fe are 8.0, 6.0, 7.0, 7.0, 6.0,
6.0, and 6.0, respectively, in units of Bohr. The dipole-dipole
interaction between slab models can be eliminated by the effective
screening medium (ESM) method \citep{Otani_First_2006}.

\section{Models}\label{models}

Bismuth surface alloys are thermodynamically stable and can be synthesized on noble metal surfaces \citep{Moreschini_Influence_2009}.
Figure 1 shows the computational model of a surface alloy composed of the
bismuth and face-centered cubic metal. We assumed here the freestanding Bi/\emph{M}
(\emph{M}=Cu, Ag, Au, Fe, Co, and Ni) surface alloy model to be the
structure where 1/3 \emph{M} atoms on the topmost atomic layer are
replaced with Bi atoms. The structure is based on Bi/Ag and
Bi/Cu(111)-(\(\sqrt{3}\times\sqrt{3}\))\emph{R}30° surface alloys of the
prior studies \citep{Ast_Giant_2007, Kaminski_Surface_2005}. We defined
the length between the Bi and \emph{M} atoms as the corrugation
parameter \(d\) (See Fig. 1(a)). We adopted the 2-atomic-layer model
(e.g.~a Bi/Ag surface alloy its underlying Ag atomic layer) for
optimizing \(d\), and 1-atomic-layer model (e.g.~only the Bi/Ag alloy)
for calculating the band structure of each freestanding surface alloy,
respectively. The 2-atomic-layer model was used for the computational model in the
previous study \citep{Bian_Origin_2013}. We used the experimental
lattice constants for each surface alloy model shown in Tab. 1. Our
calculated values of \(d\) for the Bi/Ag and Bi/Cu surface alloys are
consistent with the experimental ones
\citep{Ast_Giant_2007, Kaminski_Surface_2005}.
\begin{figure}
\begin{minipage}[t]{0.47\columnwidth}\raggedright\strut
\begin{enumerate}
\def\labelenumi{(\alph{enumi})}
\item
\end{enumerate}

\includegraphics[width=\columnwidth]{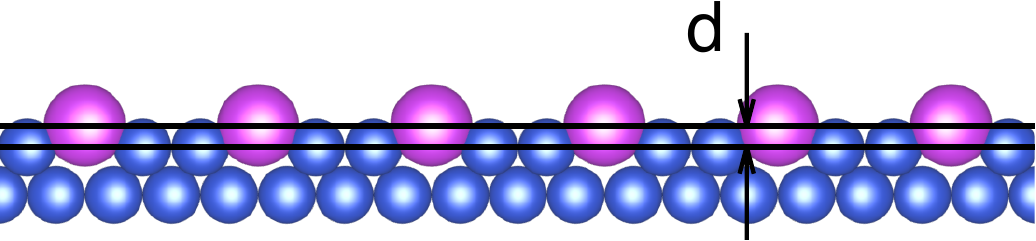}\\
\strut\end{minipage}
\begin{minipage}[t]{0.47\columnwidth}\raggedright\strut
\begin{enumerate}
\def\labelenumi{(\alph{enumi})}
\setcounter{enumi}{1}
\item
\end{enumerate}

\includegraphics[width=\columnwidth]{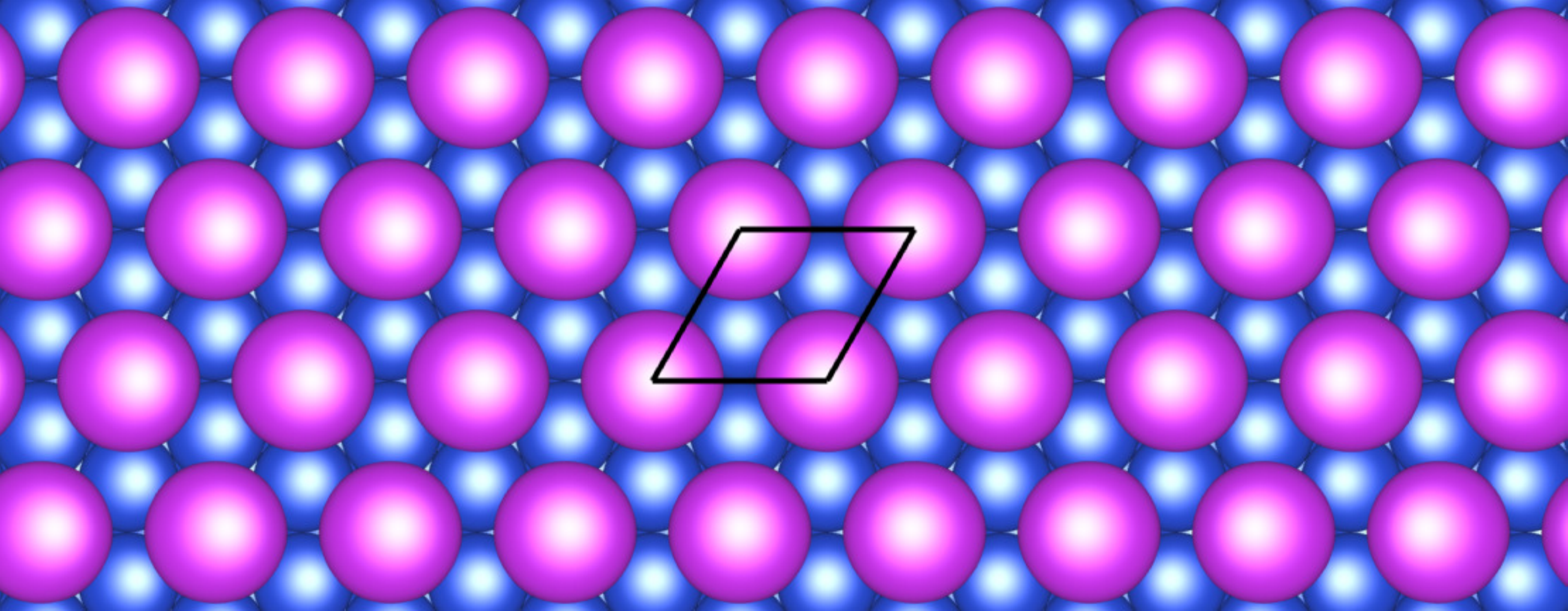}\\
\strut\end{minipage}\tabularnewline
\caption{Atomic structure of the Bi/\emph{M} surface alloy. (a) Side
view. \(d\) denotes the corrugation parameter. (b) Top view. the rhombus
represents the unit cell. The in-plane cell length is
\((\sqrt{6}/2)a_{M}\), where \(a_M\) is the lattice constant for
\emph{M}(100).}\tabularnewline
\end{figure}

\begin{table*}
\centering
\caption{Lattice constants \(a_{M}\) for \emph{M}(100) and optimum
corrugation parameters \(d\) for each surface alloy model
Bi/\emph{M}.}\tabularnewline
\begin{tabular}{lcccccc}
\toprule
Bi/\emph{M} & Bi/Cu & Bi/Ag & Bi/Au & Bi/Ni & Bi/Co & Bi/Fe\tabularnewline
\midrule
\(a_{M}\)(Å, expt.) & 3.615\citep{Ingen_Laser_1994} &
4.1\citep{Ingen_Laser_1994} & 4.0773\citep{Ellner_On_1991} &
3.53\citep{Cable_Magnetic_1995} & 3.5656\citep{Singh_High_1985} &
3.613\citep{Schlosser_Calculation_1973}\tabularnewline
\(d\)(Å) & 1.015 & 0.690 & 0.785 & 0.912 & 0.929 & 1.043\tabularnewline
\(d\)(Å, expt.) & 1.02\citep{Kaminski_Surface_2005} &
0.65\citep{Gierz_Structural_2010} & - & - & - & -\tabularnewline
\bottomrule
\end{tabular}
\end{table*}
\section{Results and Discussion}\label{results-and-discussion}

Figure 2(a)-(c) shows the band structures for the freestanding surface
alloys composed of bismuth and noble metals (Cu, Ag, Au). There are two
notable Rashba splittings of free-electron-like bands around \(\Gamma\)-point for
each surface alloy. The Rashba bands are 3/4 filled with valence
electrons. We focused on the both splittings in the band structures, and
evaluated the Rashba paramters: \(k_R\), \(E_R\), and \(\alpha_R\) (See
Tab. 2).

In the case of the Bi/Ag surface alloy, our calculated values for the
lower Rashba splitting in the Bi/Ag surface alloy (\(k_R\)=0.124
Å\textsuperscript{-1}; \(E_R\)=0.177 eV; \(\alpha_R\)=2.85 eV\(\cdot\)Å)
are agreement with experimental ones (\(k_R\)=0.13
Å\textsuperscript{-1}; \(E_R\)=0.2 eV; \(\alpha_R\)=3.05 eV\(\cdot\)Å
\citep{Ast_Giant_2007}). Besides, Our result that the Rashba momentum
offset \(k_R\) is the value of 0.124 Å\textsuperscript{-1} for the lower
splliting, and 0.075 Å\textsuperscript{-1} for the upper, respectively,
are also consistent with those of the previous theoretical study
\citep{Bian_Origin_2013}: \(k_R\)=0.144 Å\textsuperscript{-1} at
\(d\)=0.65 Å for the lower, and \(k_R\)=0.09 Å\textsuperscript{-1} at
\(d\)=0.8 Å for the upper.

For the Bi/Cu alloy, the Rashba coefficient \(\alpha_R\) of the Bi/Cu is
comparable to that of the Bi/Ag. However, our results for the Bi/Cu differ
from the earlier studies, where the Rashba parameters of \(k_R\)=0.03
Å\textsuperscript{-1}, \(E_R\)=0.015 eV, and \(\alpha_R\)=1 eV\(\cdot\)Å
were reported \citep{Moreschini_Influence_2009}. That difference might
be caused by the thickness of Cu in the computational model, for our calculated values for the 10-atomic-layer model are \(k_R\)=0.036 Å\textsuperscript{-1}, \(E_R\)=0.015 eV, and \(\alpha_R\)=0.83 eV\(\cdot\)Å. For the
Bi/Au alloy, the Rashba coefficient \(\alpha_R\) is slightly smaller
compared with the other noble metal alloys.
\begin{figure*}
\begin{minipage}[t]{0.30\textwidth}\raggedright\strut
\begin{enumerate}
\def\labelenumi{(\alph{enumi})}
\item
\end{enumerate}

\includegraphics[width=\columnwidth]{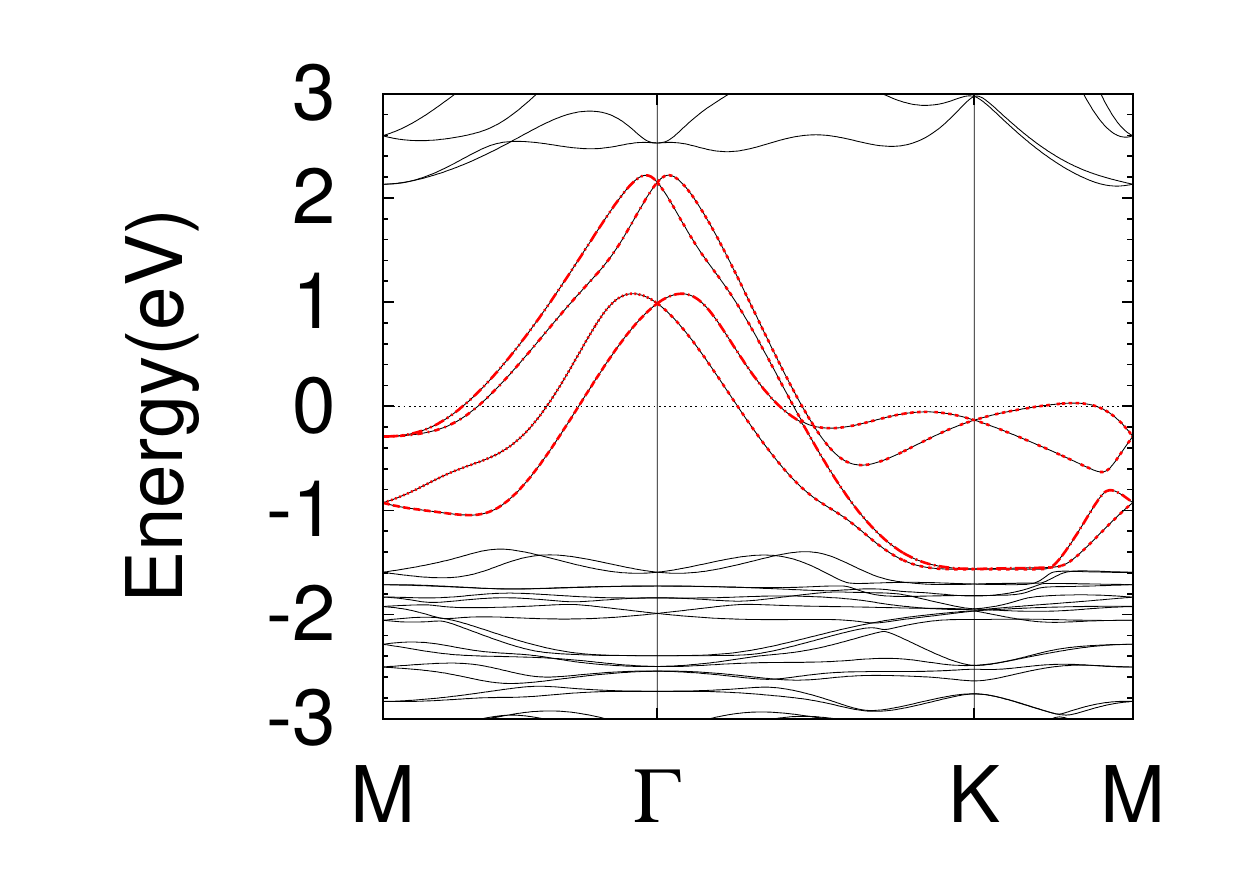}\\
\strut\end{minipage}
\begin{minipage}[t]{0.30\textwidth}\raggedright\strut
\begin{enumerate}
\def\labelenumi{(\alph{enumi})}
\setcounter{enumi}{1}
\item
\end{enumerate}

\includegraphics[width=\columnwidth]{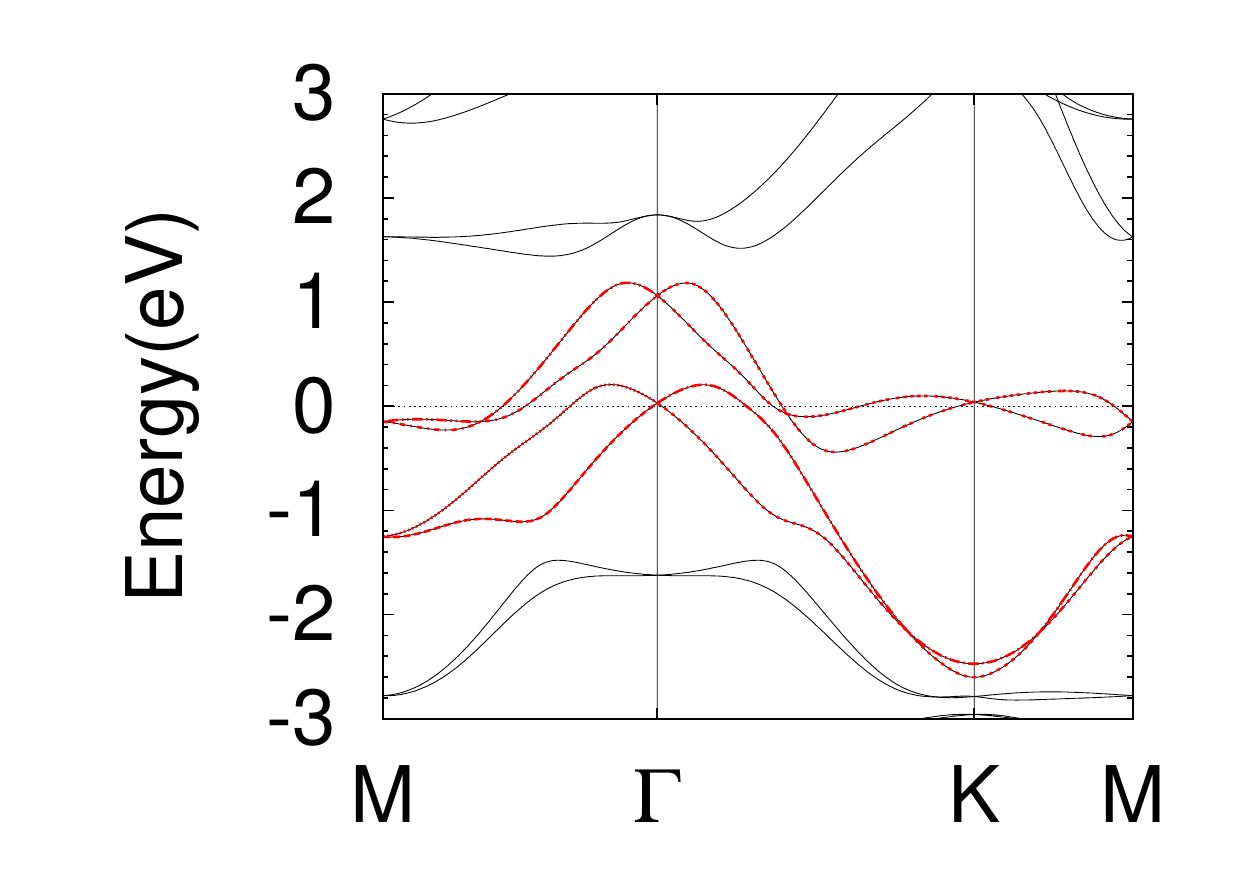}\\
\strut\end{minipage}
\begin{minipage}[t]{0.30\textwidth}\raggedright\strut
\begin{enumerate}
\def\labelenumi{(\alph{enumi})}
\setcounter{enumi}{2}
\item
\end{enumerate}

\includegraphics[width=\columnwidth]{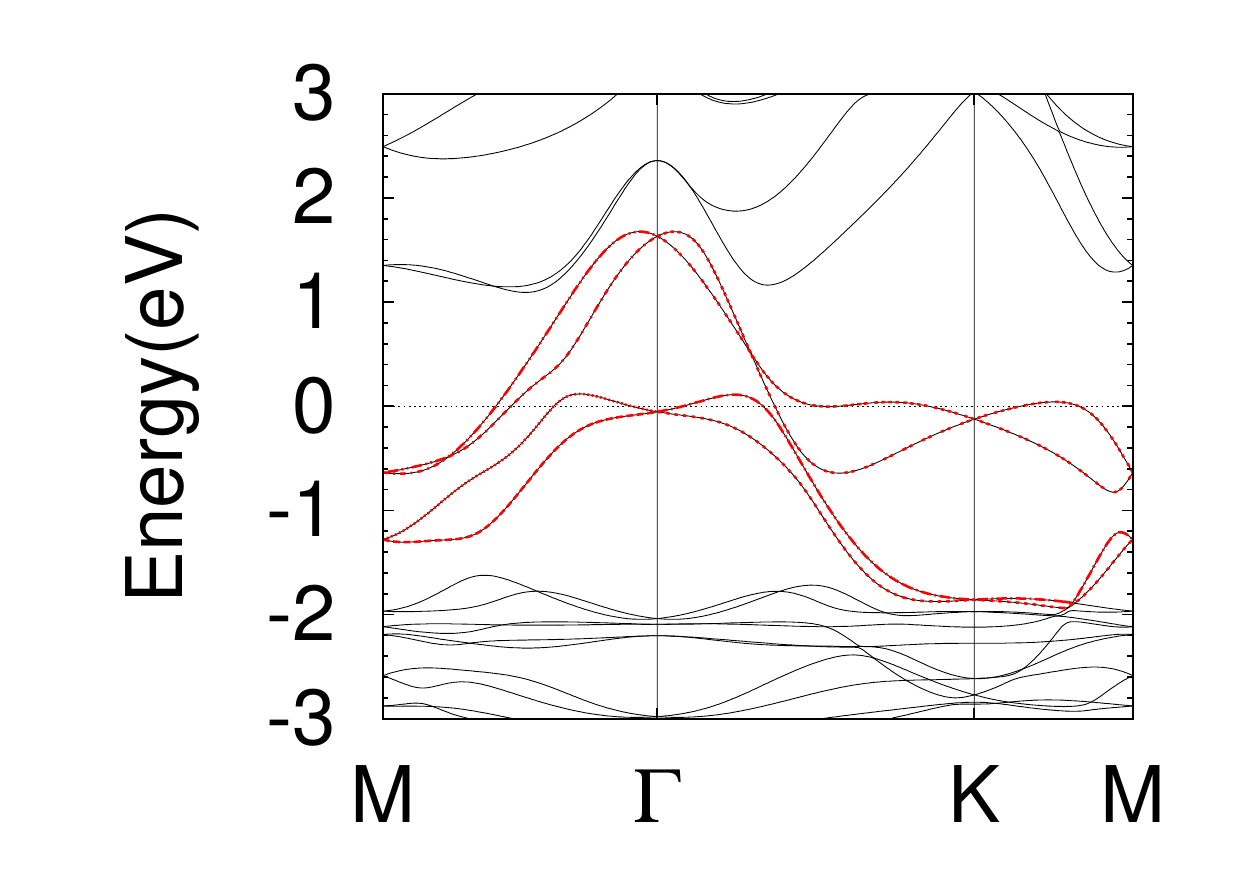}\\
\strut\end{minipage}\tabularnewline
\caption{Band structures for each freestanding surface alloy: (a) Bi/Cu; (b)
Bi/Ag; (c) Bi/Au.}\tabularnewline
\end{figure*}
There is a trend in the order of magnitude of \(\alpha_R\): Bi/Ag
\textgreater{} Bi/Cu \textgreater{} Bi/Au (the lower splitting);
Bi/Cu \textgreater{} Bi/Ag \textgreater{} Bi/Au (the upper
splitting). Since the lower splitting is larger than the upper one in the
momentum offset \(k_R\), the lower is seemed to be more important to
transport properties if the Fermi level lies around it.

We also calculated the Bi/Ni, Bi/Co, and Bi/Fe surface alloys under
non-magnetic condition. Figure 3(a)-(c) shows the band structures for
the freestanding surface alloys composed of bismuth and 3\emph{d}-transition metals
(Ni, Co, Fe). As with the surface alloys composed of Bi and noble
metals, there are two Rashba splittings for each surface alloys. As for
the transition metals, there is a trend in Rashba parameters for both splitings: Bi/Ni \(>\) Bi/Co
\(>\) Bi/Fe (See
Tab. 2).

\begin{figure*}
\begin{minipage}[t]{0.30\textwidth}\raggedright\strut
\begin{enumerate}
\def\labelenumi{(\alph{enumi})}
\item
\end{enumerate}

\includegraphics[width=\columnwidth]{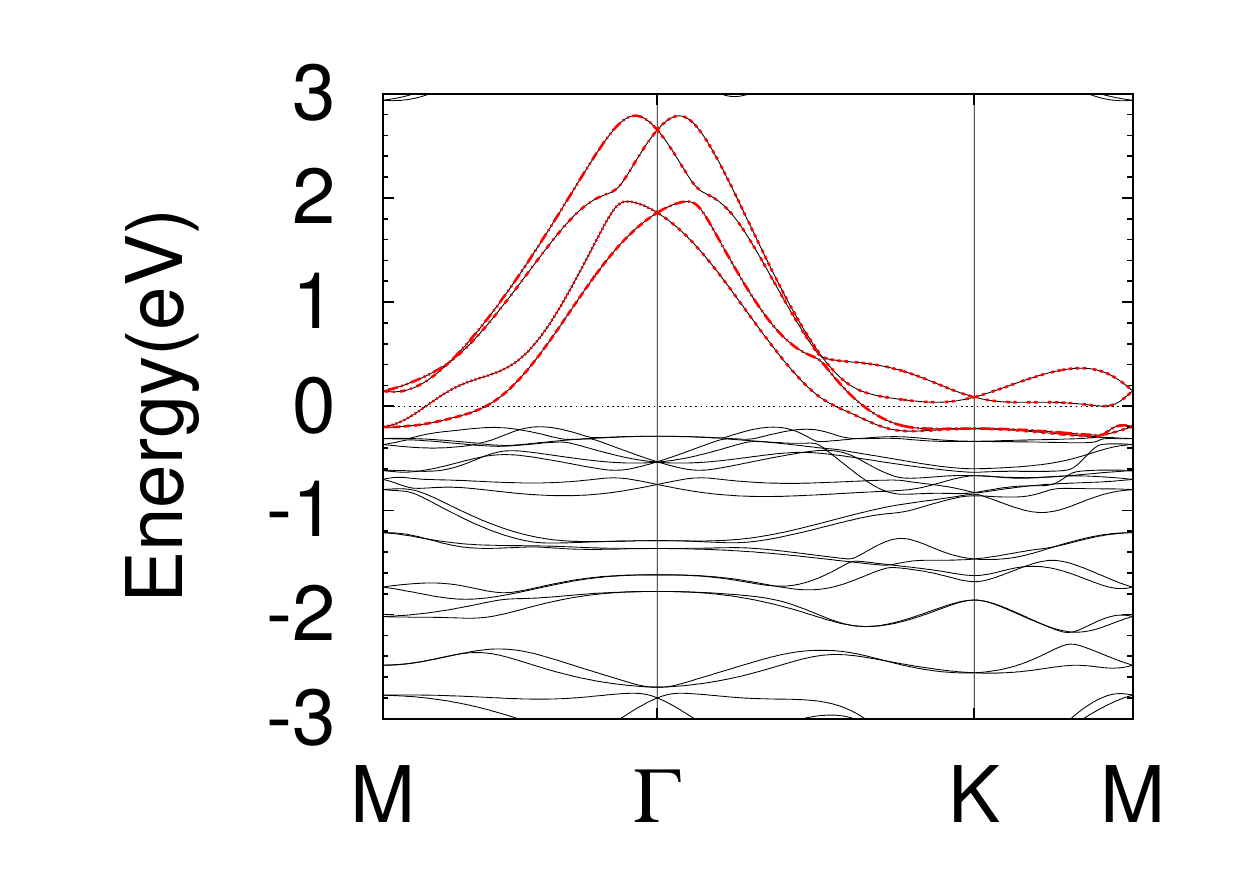}\\
\strut\end{minipage}
\begin{minipage}[t]{0.30\textwidth}\raggedright\strut
\begin{enumerate}
\def\labelenumi{(\alph{enumi})}
\setcounter{enumi}{1}
\item
\end{enumerate}

\includegraphics[width=\columnwidth]{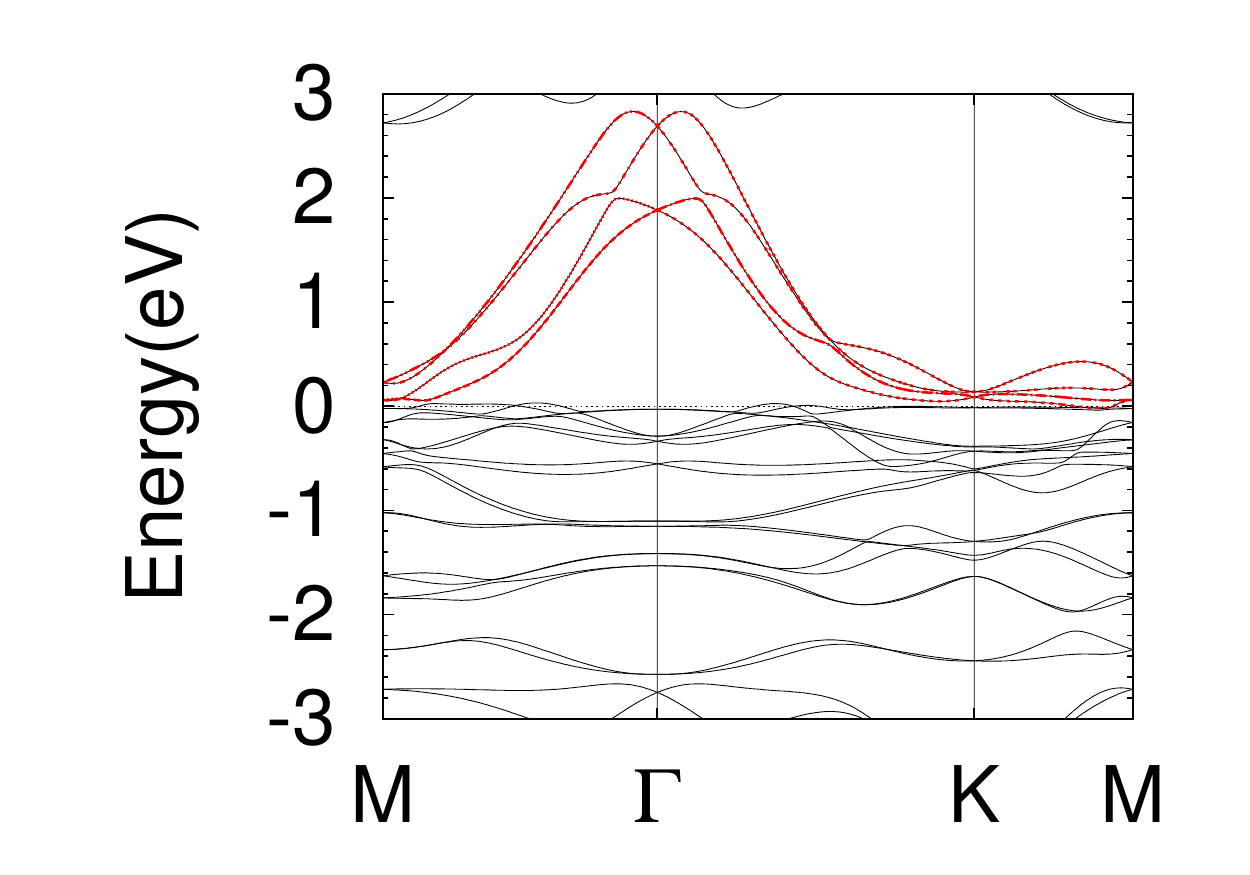}\\
\strut\end{minipage}
\begin{minipage}[t]{0.30\textwidth}\raggedright\strut
\begin{enumerate}
\def\labelenumi{(\alph{enumi})}
\setcounter{enumi}{2}
\item
\end{enumerate}

\includegraphics[width=\columnwidth]{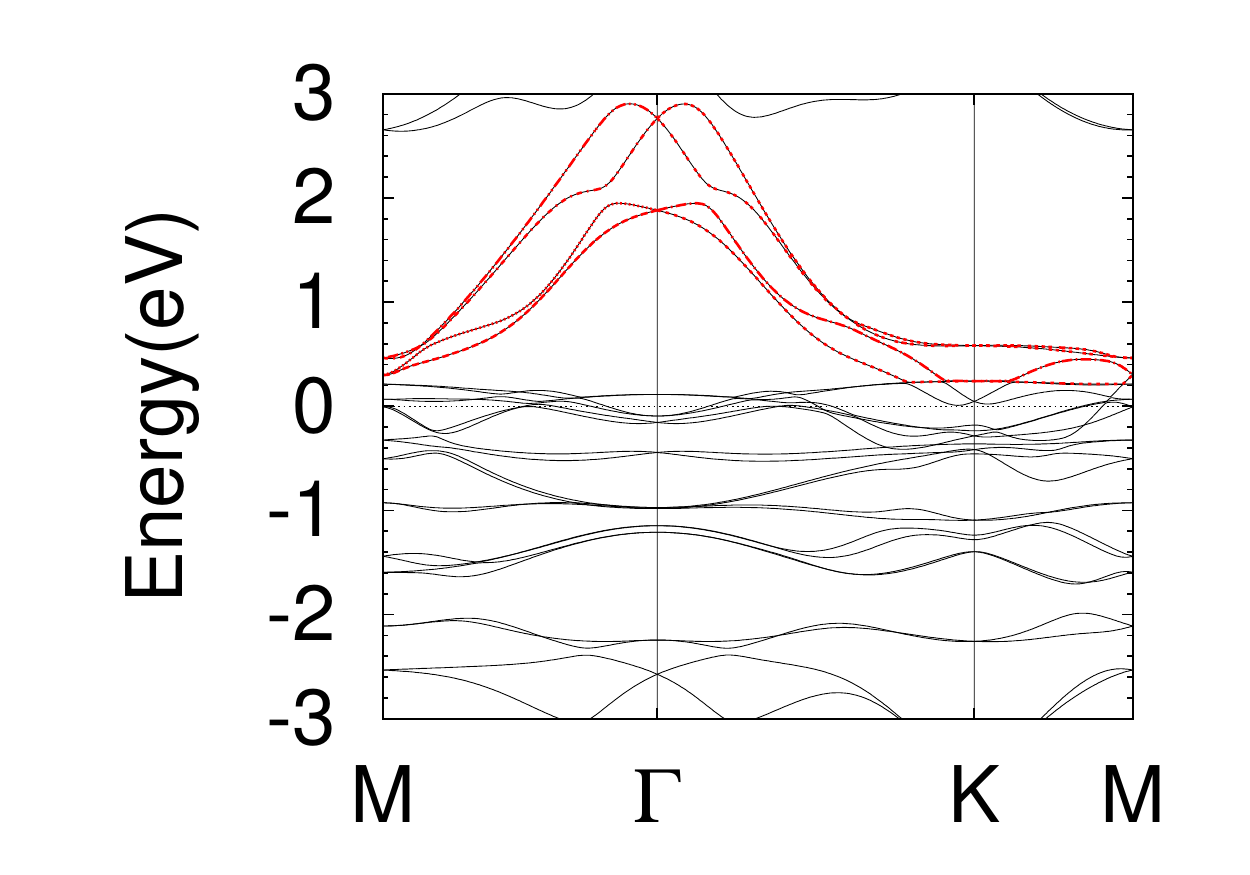}\\
\strut\end{minipage}\tabularnewline
\caption{Band structures for each freestanding surface alloy: (a) Bi/Ni; (b)
Bi/Co; (c) Bi/Fe.}\tabularnewline
\end{figure*}
\begin{table*}
\centering
\caption{Rashba parameters in the Bi/\emph{M} alloys: $\alpha_R$ is the Rashba coefficient, $E_R$ the Rashba energy, $k_R$ the Rashba momentum offset.}\tabularnewline
\begin{tabular}{lcccccc}
\toprule
\emph{M} & Cu & Ag & Au & Ni & Co & Fe\tabularnewline
\midrule
(Upper splitting) & & & & & &\tabularnewline
\(k_R\)(Å\textsuperscript{-1}) & 0.036 & 0.075 & 0.046 & 0.067 & 0.077 &
0.082\tabularnewline
\(E_R\)(eV) & 0.068 & 0.123 & 0.044 & 0.135 & 0.140 &
0.139\tabularnewline
\(\alpha_R\)(eV\(\cdot\)Å) & 3.76 & 3.28 & 1.91 & 4.05 & 3.71 &
3.40\tabularnewline
(Lower splitting) & & & & & &\tabularnewline
\(k_R\)(Å\textsuperscript{-1}) & 0.072 & 0.124 & 0.206 & 0.094 & 0.113 &
0.115\tabularnewline
\(E_R\)(eV) & 0.093 & 0.177 & 0.168 & 0.107 & 0.112 &
0.067\tabularnewline
\(\alpha_R\)(eV\(\cdot\)Å) & 2.59 & 2.85 & 1.63 & 2.28 & 1.98 &
1.17\tabularnewline
\bottomrule
\end{tabular}
\end{table*}

The Rashba coefficient \(\alpha_R\) are thought to originate from the asymmetry of the surface state (SS) \citep{Nagano_A_2009, Bentmann_Spin_2011}.
The mechanism of the Rashba-Bychkov effect at the surface due to the asymmetric SS was explained using an expression: \(\alpha_R=(2/c^2)\int d\vec{r}\partial_zV|\psi_{SS}|^2\), where \(c\) is the speed of light, \(V\) the potential, and \(\psi_{SS}\) the wave function of SS \citep{Nagano_A_2009}.
As mentioned above, in the case of the Bi/Cu alloy, our calculated \(\alpha_R\) for 1- and 10-atomic-layer model are 2.59 and 0.83 eV\(\cdot\)Å, respectively, and thus, the magnitude of \(\alpha_R\) is strongly dependent on the thickness of Cu.
However, for the Bi/Ag alloy, \(\alpha_R\) is insensitive to the thickness of Ag (2.85 eV\(\cdot\)Å (1-atomic-layer); 2.82 eV\(\cdot\)Å (10-atomic-layer)).
This difference between the Bi/Cu and Bi/Ag may be due to the difference in the degree of the localization of SS, and in terms of the thickness dependence of \emph{M}, its localization is seemed to be stronger for the Bi/Ag than Bi/Cu.
Indeed, it was reported that the strong localized SS exists for the Bi/Ag alloy \citep{Bian_Origin_2013, Moreschini_Influence_2009}.
For all the 1-atomic-layer model, \(\alpha_R\) is so large as to comparable with the Bi/Ag surface alloy, with the thinness of the surface alloy films confining the wave functions including the SS.
In other words, the strong localization of the SS may enhance the asymmetry of the charge distribution for the SS, which makes \(\alpha_R\) large.
The localization of the SS as well as the corrugation parameter \(d\) may be important to \(\alpha_R\).

\section{Conclusion}\label{conclusion}

We performed density functional calculations for
ultra-thin bismuth surface alloys: Bi/\emph{M}(111)-(\(\sqrt{3}\times\sqrt{3}\))\emph{R}30°, and calculated
Rashba parameters to predict its trend for two notable Rashba splittings of free-electron-like bands around \(\Gamma\)-point. For noble metals, we found a trend in the Rashba coefficents \(\alpha_R\): Bi/Ag
\(>\) Bi/Cu \(>\) Bi/Au (the lower splitting);
Bi/Cu \(>\) Bi/Ag \(>\) Bi/Au (the upper
splitting).
The lower values may be important to
transport properties because the lower is larger in the
momentum offset \(k_R\).
As for the transition metals, there is a trend in \(\alpha_R\): Bi/Ni
\(>\) Bi/Co \(>\) Bi/Fe.
Not only the corrugation parameter \(d\) but the localization of the surface states may be important to \(\alpha_R\).
Our finding may lead to design efficient
spin-charge conversion materials.

\section{Acknowledgements}\label{acknowledgements}

The authors thank K. Kondou for invaluable discussions.
This work was supported by Grant-in-Aid for Scientific Research on Innovative Area, ”Nano Spin Conversion Science” (Grant No.15H01015).
The work was partially supported by Grants-in-Aid on Scientific Research
under Grant Nos. 25790007 and 16K04875 from Japan Society for
the Promotion of Science. This work was also supported in part by MEXT as a
social and scientific priority issue (Creation of new functional devices
and high-performance materials to support next-generation industries) to
be tackled by using post-K computer. Computations in this research were
also performed using supercomputers at ISSP.


\bibliography{readcube_export.bib}

\end{document}